\documentstyle[preprint,revtex]{aps}

\def\llrr{\leftrightarrow}
\def\eqn#1{(\ref{#1})}
\font\sml=cmr10 scaled\magstep1
 \def\beg{\begin{equation}}
 \def\eb{\end{equation}}
 \def\<{\noindent }
   \def\z#1#2{\ifcase#1 {\overline {#2}} \or                   %\z0=overline
   {\null\ifmmode{\underline #2}\else{\underbar #2}\fi} \or    %\z1=underline
   {\if #2i {\hat\imath } \else\if #2j {\hat\jmath }           %\z2=hat
      \else {\hat {#2}} \fi\fi} \or
   {\if #2i {\vec\imath} \else\if #2j {\vec\jmath}             %\z3=vector
      \else {\vec #2} \fi\fi} \or
   {\if #2i {\tilde\imath} \else\if #2j {\tilde\jmath}         %\z4=twidle
       \else {\tilde #2} \fi\fi} \or
   {{\text{\b{$#2$}}}} \or   % Short underscore in math mode.  %\z5s
   {{\bf #2}} \fi}

 \def\za{{{}\ifmmode\alpha\else$\alpha$ \fi}}
 \def\zb{{{}\ifmmode\beta\else$\beta$ \fi}}
 \def\zc{{{}\ifmmode\psi\else$\psi$ \fi}}
 \def\zC{{{}\ifmmode\Psi\else$\Psi$ \fi}}
 \def\zd{{{}\ifmmode\delta\else$\delta$ \fi}}
 \def\zD{{{}\ifmmode\Delta\else$\Delta$ \fi}}
 \def\ze{{{}\ifmmode\epsilon\else$\epsilon$ \fi}}
 \def\zf{{{}\ifmmode\phi\else$\phi$ \fi}}
 \def\zF{{{}\ifmmode\Phi\else$\Phi$ \fi}}
 \def\zg{{{}\ifmmode\gamma\else$\gamma$ \fi}}
 \def\zG{{{}\ifmmode\Gamma\else$\Gamma$ \fi}}
 \def\zh{{{}\ifmmode\eta\else$\eta$ \fi}}
 \def\zi{{{}\ifmmode\iota\else$\iota$ \fi}}
 \def\zI{{{}\ifmmode\infty\else$\infty$ \fi}}        %
 \def\zk{{{}\ifmmode\kappa\else$\kappa$ \fi}}        %
 \def\zl{{{}{}\ifmmode\lambda\else$\lambda$ \fi}}
 \def\zL{{{}\ifmmode\Lambda\else$\Lambda$ \fi}}
 \def\zm{{{}\ifmmode\mu\else$\mu$ \fi}}
 \def\zn{{{}\ifmmode\nu\else$\nu$ \fi}}
 \def\zN{{{}\ifmmode\emptyset\else$\emptyset$ \fi}}  %
 \def\zp{{{}\ifmmode\pi\else$\pi$ \fi}}
 \def\zP{{{}\ifmmode\Pi\else$\Pi$ \fi}}
 \def\zchi{{{}\ifmmode\chi\else$\chi$ \fi}}
 \def\zr{{{}\ifmmode\rho\else$\rho$ \fi}}
 \def\zs{{{}\ifmmode\sigma\else$\sigma$ \fi}}
 \def\zS{{{}\ifmmode\Sigma\else$\Sigma$ \fi}}
 \def\zt{{{}\ifmmode\tau\else$\tau$ \fi}}
 \def\zu{{{}\ifmmode\upsilon\else$\upsilon$ \fi}}
 \def\zU{{{}\ifmmode\Upsilon\else$\Upsilon$ \fi}}
 \def\zv{\partial}                                   %
 \def\zw{{{}\ifmmode\omega\else$\omega$ \fi}}
 \def\zW{{{}\ifmmode\Omega\else$\Omega$ \fi}}
 \def\zx{{{}\ifmmode\xi\else$\xi$ \fi}}
 \def\zX{{{}\ifmmode\Xi\else$\Xi$ \fi}}
 \def\zy{{{}\ifmmode\theta\else$\theta$ \fi}}
 \def\zY{{{}\ifmmode\Theta\else$\Theta$ \fi}}             %\zZ1=display
 \def\zz{{{}\ifmmode\zeta\else$\zeta$ \fi}}               %\zZ2=text
 \def\zZ#1{\ifcase#1 {}\or \displaystyle \or \textstyle   %\zZ3=sub or super
        \or \scriptstyle \or \scriptscriptstyle \fi}      %\zZ4=scriptscript
 
 \def\half{{1\over2}}
 \def\third{{1\over3}}
 \def\fourth{{1\over4}}
 \def\ahalf#1{{#1\over2}}
 
\begin{document}
%%%%%%%%%%%%%%%%%%%%%%%%%%%%%%%%%%%%%%%%%%%%%%%%%%%%%%%%%%%%%%%%%%%%%%%%%%%%%%%
\preprint{WISC-MIL-92-TH-15}
\begin{title}
\Large Quark Neutron Layer Stars
\footnote{To Appear in The Astrophysical Journal }
\end{title}
\bigskip
\centerline{\large \quad Philip A. Carinhas
\footnote{email: carinhas@csd4.csd.uwm.edu}
}
\begin{instit}
\sml
Department of Physics \\
University of Wisconsin \\
Milwaukee, Wisconsin 53201
\end{instit}
% \receipt{1 January 1989}
\begin{abstract}
%%%%%%%%%%%%%%%%%%%%%%%%%%%%%%%%%%%%%%%%%%%%%%%%%%%%%%%%%%%%%%%%%%%%%%%%%%%%%
\baselineskip=20pt
{
Typical nuclear equations of state and a quark bag model, surprisingly,
allow compact stars with alternate layers of neutrons and quarks.
One can determine on the basis of the Gibbs free energy which phase,
nuclear or quark, is energetically favorable.  Using the nuclear equation
of state of Wiringa, and a quark equation of state given by
Freedman and McLerran, the allowed quark parameter space for such layer
stars is searched.
%%%%%%%%%%%%%%%%%%%%%%%%%%%%%%%%%%%%%%%%%%%%%%%%%%%%%%%%%%%%%%%%%%%%%%%%%%%%%
This paper differs from past work in that configurations are found in which
quark matter is located exterior and interior to shells of nuclear matter,
i.e., dependent on quark parameters, a star may contain several alternating
layers of quark and nuclear matter.
%%%%%%%%%%%%%%%%%%%%%%%%%%%%%%%%%%%%%%%%%%%%%%%%%%%%%%%%%%%%%%%%%%%%%%%%%%%%%
Given the uncertainty in the quark parameter space, one can estimate the
probability for finding pure neutron stars, pure quark stars (strange stars),
stars with a quark core and a nucleon exterior, or layer stars.
Several layer models are presented. The physical characteristics,
stability, and results of a thorough search of the quark parameter space
are presented.
%%%%%%%%%%%%%%%%%%%%%%%%%%%%%%%%%%%%%%%%%%%%%%%%%%%%%%%%%%%%%%%%%%%%%%%%%%%%%
}

\vskip32pt
\noindent astro-ph/9304021 \hfill

\end{abstract}

\pagebreak
%%%%%%%%%%%%%%%%%%%%%%%%%%%%%%%%%%%%%%%%%%%%%%%%%%%%%%%%%%%%%%%%%%%%%%%%%%%%%
\baselineskip=20pt
%%%%%%%%%%%%%%%%%%%%%%%%%%%%%%%%%%%%%%%%%%%%%%%%%%%%%%%%%%%%%%%%%%%%%%%%%%%%%
\section{Introduction}
Although thought to be composed of neutrons, there is still no compelling
evidence for the microscopic description of matter in compact stars, partly
because of the difficulty in experimentation at these extreme densities.

Recently, there has been a great interest in QCD phase transitions and the
composition of compact stars, including quark stars, strange stars, neutron
stars with pion condensate, gluon condensate, and quarks
\cite{benvenuto91,campbell90,ellis91,grassi,haensel86}.

This paper investigates the possibility that dense stars with more than one
layer of neutron matter and quark matter exist.
If the Witten hypothesis is correct, the ground state of matter may
be strange and so dense stars with these structures could occur in nature
\cite{witten84}.

The Wiringa \cite{wiringa} equation of state and a quark equation of state
given by Freedman and McLerran are used \cite{freedman78}.
Given this and the possible quark parameter range, stars which
have several layers of alternate neutron and quark matter are constructed.
For a maximum and minimum pressure, the quark parameter space is classified
according to the most complex structure that can be constructed from these
parameters. The fractional volumes of the quark parameter space for each
stellar type are calculated.  Given the uncertainty in the quark parameters,
there is a finite area in the quark parameter space for which layer stars
can exist.

%%%%%%%%%%%%%%%%%%%%%%%%%%%%%%%%%%%%%%%%%%%%%%%%%%%%%%%%%%%%%%%%%%%%%%%%%%
This work assumes two phases of matter that are each electrically neutral.
This leads to a discontinuous quark-neutron phase transition, with
discontinuities in the energy and mass densities.
%%%%%%%%%%%%%%%%%%%%%%%%%%%%%%%%%%%%%%%%%%%%%%%%%%%%%%%%%%%%%%%%%%%%%%%%%%
Glendenning (1992) has recently argued that if one requires only overall
neutrality, allowing two intermixed phases to each have a net charge,
the phase transition from nucleon matter to quark matter can occur gradually
over a finite range of pressure \cite{glendenning92}. The energy density and
chemical potentials would be continuous through the transition region.
Although Glendenning has neglected surface effects of the two embedded phases,
the general features of the mixed phase described there should not change.
Note that some of the phase transitions which occur in the current work
(that are close in pressure) would likely be part of a single mixed phase
region under Glendenning's assumptions.

\section{Quark Parameters and the Bag Model}
\label{sec:QPBM}

 In this section, a version of the bag model due to Freedman and McLeran
\cite{freedman78} is introduced and the relevant bounds on the quark parameter
space is discussed. Strange matter is modeled as a degenerate Fermi gas of
$u$, $d$, and $s$ quarks and electrons (positrons). Chemical equilibrium
between the three flavors is maintained by the reactions

  \beg
  \begin{array}{ccc}
 d      \; &\to   \;&\;  u + e + \z0\zn_e  \cr
 u + e\;\; &\to   \;&    d + \zn_e         \cr
 s      \; &\to   \;&\;  u + e + \z0\zn_e  \cr
 s + u\;\; &\llrr \;&    d + u             \cr
 \label{react}
\end{array}
\eb

The chemical potentials of the lost neutrinos are set to zero since they
interact very weakly. It is assumed that the strange quark has a mass
$m_s=m$, and no gluon exchange is considered. The above reactions imply that

\beg\zm_d = \zm_s \equiv \zm \quad ; \quad \zm_u + \zm_e = \zm.\eb

\<Overall charge neutrality for bulk matter implies

  \beg 2n_u - n_d - n_s - 3n_e = 0 \label{neutral}\eb

\<where the number densities are given in terms of the thermodynamic potentials
$\zW_a$ ($a=u,d,s,e$) are given in the appendix,

 \beg n_a = - {\zv \zW_a \over \zv \zm_a } \label{defn} \eb

\<A consequence of \eqn{react} and \eqn{neutral} is that only one chemical
potential is left independent. The energy density of the fermions is
$\sum_a(\zW_a+n_a\zm_a)$. The vacuum associated with bulk quark matter is
also assumed to contain a positive energy per volume, called the bag energy,
denoted by $B$. The total energy density is therefore,

\beg \ze = \sum_a(\zW_a + n_a\zm_a) + B \label{epsi} \eb

The quark parameters are then given by $\za_c$, the strong coupling constant
of QCD, $m_s$, and the bag constant $B$. There is also another parameter,
the renormalization point $\zr_r$ which is given the constant value of
$\zr_r=313MeV$. See \cite{alcock86} for a discussion of this value.
There are effectively 3 parameters to consider.

\<The baryon number density is given by $n_A=\third(n_u + n_d + n_s)$.
Chemical equilibrium between the Fermi pressure and vacuum pressure is achieved
via

\beg {\zv \over \zv \zm }({\ze/n_A})=0 \label{equil1}\eb

\<Equivalently,

\beg\sum_a\zW_a = -B \label{equil2}\eb

\<For known values of $B$ and $m$, one can solve for $\zm$ in \eqn{equil1},
yielding equilibrium number densities and energy densities, and the energy
per baryon. Curves of constant $E/A$ are calculated (numerically) and
determine the value of $B^\fourth$ for which non-strange quark matter has
an energy per baryon of $930 +\zD \; MeV$.
To the left of this line, nuclei with high atomic number are unstable
against decay into nonstrange quark matter. \zD, determined in \cite{farhi},
is due to shell effects in nuclei, and the value $930$ is the energy per baryon
for the most stable nuclei, iron. The allowed values of $B^\fourth$ narrow
as M increases as can be seen in Fig.\ \ref{eoa}.
This places lower limits on $B$ as a function of $\za_c$.

\firstfigfalse
\vbox{
\vskip -.65truein
\figure{\baselineskip=14pt
Curves of constant {\scriptsize $E/A=934\; MeV$}. The value of $\alpha_c$ is
given by the label at the top of each curve. The dotted lines indicate the
minimum value, $B_{min}$, allowed.
\label{eoa}}
     }

\<There is no similar arguement for an upper limit on $B$, or the other
quark parameters.

  Since we are using a perturbative expansion in powers of $\za_c$,
one must be skeptical of the quark equation of state in a regime where
$\za_c\geq 1$. For $\za_c > 1.0$, second order contributions in $\za_c^2$
are greater than the zero order result and the energy density is unbounded
from below as $T \to \infty$ \cite{chin78}. Also, the typical energy scale
of neutrons stars is well below that of deconfinement where the bag model
is appropriate. Dispite these considerations, results up to $\za_c = 1.5$
(where the equation of state becomes unstable) are included. Some limits on
values for the quark parameters are given by several authors below.
The first four references come from fits to heavy ion experiments.
The last values are those used by the author.

\vskip30pt

\vbox{
\begin{quasitable}
\caption{Range of Quark Parameters.}
\centerline{\bf Quark Parameters for the Bag Model.}
\begin{tabular}{c||ccc}
\tableline
Reference& $\za_c$ & $m_s (MeV)$ &$B^\fourth (MeV)$ \\
\tableline
\cite{benvenuto91} &$.5-.6    $&$150-200 $&$ 145           $\\
\cite{degrand}     &$2.2      $&$280     $&$ 145           $\\
\cite{bart}        &$2.0      $&$283     $&$ 149           $\\
\cite{chan}        &$2.8      $&$340     $&$ 120           $\\
\cite{carlson}     &$\leq 1.0 $&$288     $&$ \sim 200-220  $\\
USED               &$ 0-1.5   $&$0 - 330 $&$  B_{min}-240  $\\
\end{tabular}
\end{quasitable}
     }
%%%%%%%%%%%%%%%%%%%%%%%%%%%%%%%%%%%%%%%%%%%%%%%%%%%%%%%%%%%%%%%%%%%%%%%%%%%%%%%
\section{Phase Transitions and Layer Stars }
 This section discusses the construction and physical characteristics of
layer stars. The matter type is determined by the Gibbs free energy criterion
for two competing types of matter. The structure of layer stars is examined
and compared to known stellar models.

 Consider the Wiringa equation of state and the quark bag equation of
state, assuming, for the moment, that quark matter is stable at the core.
The Oppenheimer Volkoff (OV) equations  governing hydrodynamic equilibrium are

\beg
{\large
\begin{array}{ccc}
{dm\over dr}\; \;  &=& 4\pi r^2 \ze                                      \\
{dP\over dr}\; \;  &=& -{1\over r}{(P + \ze)(m + 4\pi Pr^3)\over (r-2m)} \\
\end{array}
}
\label{tov}
\eb

\<where $m$ is the mass interior to radius $r$, $P$ is pressure, and $r$
is the radial co-ordinate. Numerical integration of these equations for
spherical stars is outlined in \cite{shapiro}.

At each step of integration, calculate the Gibbs free energy per baryon,

 \beg g = { \ze + P \over n} \eb

\<(It is assumed that $T$ is small compared to the Fermi energy of the matter
and thus $T$ and  $s$ can be neglected.)
Label quark matter as Q and nuclear matter as N.
When the Gibbs free energy of quark matter surpasses (for the same pressure)
the Gibbs free energy of nuclear matter, the equation of state (or table) is
changed to that of nuclear matter.
This point represents the phase transition between the two materials.
The OV equations are then integrated with the new equation of state
until the pressure drops to zero and the program terminates.
The number and hence the energy density are not continuous at the phase
transition, however, $g$ and $P$ are continuous everywhere. The hybrid
case has been studied for first and second order phase transitions using
a mean field equation of state \cite{ellis91}.

    If the quark parameters are such that the quark equation of state
crosses the neutron equation with the neutron equation of state
having lower Gibbs free energy at low pressure, nucleons will be the
stablest at low energy (at the surface of the star). If the Gibbs free
energy of the quark matter is lower at low pressure, one will have quark
matter at the surface.

  This paper considers stars which have any type of matter at the
core or surface, and have a number of layers of matter which alternate between
quark and neutron. The Gibbs free energy as a function of pressure is presented
below for four choices of quark parameters $B$, $m_s$, and $\alpha_c$ in
Fig \ref{pg}.

\vbox{
\figure{\baselineskip=14pt
Gibbs free energy vs. pressure for layer stars.
        (a) corresponds to  $\za_c =.5$,$m_s=300MeV$, $B^\fourth=140MeV$,
        (b) to     $\za_c =.5$,$m_s=220MeV$, $B^\fourth=140MeV$,
        (c) to     $\za_c =.5$,$m_s=280MeV$, $B^\fourth=132MeV$,
        (d) to     $\za_c =.5$,$m_s=230MeV$, $B^\fourth=132MeV$.
  The dotted lines correspond to the Wiringa equation of state.
  \label{pg} } }

%%%%%%%%%%%%%%%%%%%%%%%%%%%%%%%%%%%%%%%%%%%%%%%%%%%%%%%%%%%%%%%%%%%%%%%%%%%%%

For the choice of parameters, case (a) yields a star which has one phase
transition at pressure $P = 108.7\times 10^{34} dynes/cm^2$,
case (b) has three crossings at $P = .556, 2.81$, and $89.46\times
10^{34} dynes/cm^2$, cases (c) and (d) both have 2 crossings at
$4.91,97.26$, and $10.43, 86.99\times 10^{34} dynes/cm^2$ respectively.
The central pressure is that for the maximum mass Wiringa model.
%%%%%%%%%%%%%%%%%%%%%%%%%%%%%%%%%%%%%%%%%%%%%%%%%%%%%%%%%%%%%%%%%%%%%%%%%%%%%

\subsection{Quark Neutron Layer Stars}

Consider sequences based on the parameters in the previous section.
The following mass-radius and moment of inertia-radius diagrams contain
all the essential features of layer stars for both models with quark
exteriors and neutron exteriors. Note the similarity between a pure neutron
sequence and cases (a) and (b), and between pure quark sequences and
cases (c) and (d).

 \figure{\baselineskip=14pt
 Mass vs. radius for layer stars. Circles in the sequences
 correspond to phase transition from one stellar type to the next.
 The quark parameters are again given as above in Figure \ref{pg}.
 \label{rm}}

Sequence (a) has the appearance of a pure neutron mass-radius diagram,
except for the peak at the onset of the QN phase.
Also in case (b), the QN region is drawn into a smaller radius
as compared to the pure neutron part of sequence (a). This is due to the
presence of a quark core in the QN models.
For phase transition which occur at lower pressures, this feature is enhanced
to the point of meeting the pure quark sequences at low mass.
The onset of instability often coincides with a phase transition from neutron

 \figure{\baselineskip=14pt
 Moment of Inertia vs. radius for layer stars. Circles correspond to phase
 transition as above.  The quark parameters are again given as above.
 \label{ri}}

\<to quark matter. This is due to the quark equation of state being softer than
the neutron equation of state at these pressures. Figure \ref{ri} (a) most
closely resembles a pure neutron sequence. (b) has a much smaller radius at
low moment of inertia, resembling (c) and (d), as in pure quark sequences.
This is similar to the mass-radius diagrams in Fig \ref{rm}.

Accuracy for the stellar calculations were checked against constant density
star models (which are exactly solvable) to better than $3.0 \times 10^{-6}\%$
for the radius and $1.5\times 10^{-3}\%$ for the mass.

%%%%%%%%%%%%%%%%%%%%%%%%%%%%%%%%%%%%%%%%%%%%%%%%%%%%%%%%%%%%%%%%%%%%%%%%%%%%%%

\subsection{Layer Star Profiles}
The following plots show the physical parameters for four layer stars as a
function of radius of the star. In all cases, the central pressure is chosen
to be that for the maximum mass Wiringa model, $P_c=P_{max}=1.15\times 10^{34}
dynes/cm^2$.
%%%%%%%%%%%%%%%%%%%%%%%%%%%%%%%%%%%%%%%%%%%%%%%%%%%%%%%%%%%%%%%%%%%%%%%%%%%%%%
\vskip-14pt
%%%%%%%%%%%%%%%%%%%%%%%%%%%%%%%%%%%%%%%%%%%%%%%%%%%%%%%%%%%%%%%%%%%%%%%%%%%%%%
\figure{\baselineskip=14pt
Energy density-radius profile for layer stars. The discontinuities in
the plots correspond to the phase transition from one type of matter to the
next. The quark parameters are again given above in Figure \ref{pg}.
\label{re}}

\<The energy density is discontinuous at phase transitions; however, the
pressure and mass remain continuous throughout the star.

\<In (a) and (b) the matter is originally in the neutron phase and then is in
the Q phase after the first discontinuity.
There are no other transitions in (a) however, (b) undergoes 2 more transitions
to N and Q respectively. Models (c) and (d) both go through a Q, N, and Q
phase.
%%%%%%%%%%%%%%%%%%%%%%%%%%%%%%%%%%%%%%%%%%%%%%%%%%%%%%%%%%%%%%%%%%%%%%%%%%%%%%
\vskip-10pt
%%%%%%%%%%%%%%%%%%%%%%%%%%%%%%%%%%%%%%%%%%%%%%%%%%%%%%%%%%%%%%%%%%%%%%%%%%%%%%
\figure{\baselineskip=14pt
Pressure-radius profile for layer stars. Circles in the sequences
mark phase transitions from one stellar type to the next.
The quark parameters are again given above in Figure \ref{pg}.
\label{rp}}

  The pressure-radius diagrams are continuous with a flattening in
(a) and (b) at low pressure due to the stiff nuclear equation of state at
low pressure.

The mass-radius profiles are slightly more interesting. (a) and (b) are flat
near the surface, again due to the stiff nuclear equation of state
at low pressure. (c) anc (d) are almost linear in the quark phase in the
exterior region.
%%%%%%%%%%%%%%%%%%%%%%%%%%%%%%%%%%%%%%%%%%%%%%%%%%%%%%%%%%%%%%%%%%%%%%%%%%%%%%
\figure{\baselineskip=14pt
Mass-radius profiles for layer stars. Circles in the sequences mark
phase transitions from one stellar type to the next.
The quark parameters are again given above in Figure \ref{pg}.
\label{RM}}
%%%%%%%%%%%%%%%%%%%%%%%%%%%%%%%%%%%%%%%%%%%%%%%%%%%%%%%%%%%%%%%%%%%%%%%%%%%%%%
\section{Quark Parameter Space for Q-N Layer Stars}
%%%%%%%%%%%%%%%%%%%%%%%%%%%%%%%%%%%%%%%%%%%%%%%%%%%%%%%%%%%%%%%%%%%%%%%%%%%%%

  This section discusses the classification of that part of the quark parameter
space that is consistent with current knowledge of the quark parameters.
The quark parameter space is classified according to the number of phase
changes that occur between $P_{min}$ to $P_{max}$, with special regard for the
order in which the phase of matter, neutron or quark, appears.
$P_{min}$ is chosen at the pressure corresponding to the energy density where
neutron drip occurs, $\rho_{ND}\sim 4.54\times 10^{12}g/cm^3$ which corresponds
to $P_{min} \sim 7.0\times 10^{29}dyne/cm^3$ for the Wiringa equation of state.
$P_{min} = 1.0\times 10^{30}dyne/cm^3$ is taken as the more conservative value.
$P_{max}$ is taken as the pressure corresponding to the onset of
instability for the neutron regime of the Wiringa equation of state.
Associate to that point in the quark parameter space a sequence of Q's and N's,
corresponding to the matter phase of the layers emanating from the the core
of the star to the exterior. A neutron core and a quark exterior would be
labeled as a NQ for example.

\subsection{Calculation of the Paramerter Space}

\<For each point in the ($\za_c$ , $m_s$) plane of dimensions
$\zd\!\za_c \times \zd\!m_s =.01 \times 3.0 MeV$, the minimum and
maximum value of $B^\fourth$ for each stellar type is calculated to an
accuracy of $ \zD\!B = .03 MeV$.

The volume of stellar type $A$ in the quark parameter space, $V_A$, is
calculated as

\beg V_A \sim \sum_i \zD \za_{c_i} \zD m_{s_i} (B_{A+} - B_{A-})_i
             =  \zD \za_c \zD m_s \sum_i h\! B_{Ai}
\eb

\<where $h\!B_{Ai} = (B_{A+} - B_{A-})_i$ is the height of the $i_{th}$ space
that corresponds to that point in the quark parameter space,
and $\zD \za_c$, $\zD m_s$ are constants. The uncertainty in $B$ is taken
as half of the accuracy in $B$, $\zd\!B_A=\half\zD\!B $.  Given that
$\zd (B_{A+} - B_{A-})_i = 2\zd B_{Ai} = \zD B$,  the uncertainty
$\zd V_A$ is similarly calculated as

\beg \zd V_A \sim \zD \za_c \zD m_s \sum_i \zd h\!B_{Ai}
=\zD \za_c \zD m_s \sum_i \zd (B_{A+} - B_{A-})_i
=\zD \za_c \zD m_s \sum_i \zD B_{Ai} \eb

  From the results of this calculation one can determine the relative volumes
of parameter space of each stellar class.
The results are summarized in the following table.

\begin{quasitable}
\caption{Summary of Parameter Space Classification.}
\centerline{\bf Summary of Parameter Space Classification.}
\begin{tabular}{c||ll}
\tableline
Class              & $V_i/V$  &$\zD V_i/V$ \hfill \\
\tableline
N                  & 0.735    &  0.008     \hfill \\
QN                 & 0.176    &  0.004     \hfill \\
Q                  & 0.021    &  0.002     \hfill \\
NQ                 & 0.045    &  0.005     \hfill \\
QNQ                & 0.009    &  0.001     \hfill \\
NQN                & 0.006    &  0.005     \hfill \\
QNQN               & 0.006    &  0.002     \hfill \\
%NQNQN       & $\leq$ 2.0  e-4 &           \hfill \\
%QNQNQN      & $\leq$ 2.0  e-4 &           \hfill \\
%QNQNQ       & $\leq$ 2.0  e-4 &           \hfill \\
%
\end{tabular}
\end{quasitable}
%%%%%%%%%%%%%%%%%%%%%%%%%%%%%%%%%%%%%%%%%%%%%%%%%%%%%%%%%%%%%%%%%%%%%%%%%%%%%
\figure{\baselineskip=14pt
Parameter space for neutron (N) and QN stars. Slices are plotted at increments
of $\za_c=.25$.These two spaces compromise over 90\% of the total quark
parameter space. \label{wir0}}

\<N and QN occupy the majority of the entire
space (see Fig.\ \ref{wir0}). This is due in part to the monotonicity of the
equations of state, and the fact that the nuclear equation of state usually
dominates at low pressures.

The other components of the quark parameter space are much smaller
and are illustrated below. The first two are next largest spaces, NQN and NQ,
which together occupy about 4.4\% of the quark parameter space.

\figure{Parameter space for NQN and NQ stars. \label{wir11}}

\<The NQN space occurs precisely above in $B^\fourth$ the NQ space as a
result of phase transitions at low pressure (density). NQN is a very thin
space as can be seen in Fig \ref{wir11}. This is due to two phase
transitions which occur at low pressure.

\<The next two spaces, QNQ and QNQN, occupy about 1.3\% of the quark parameter
space. The QNQN space lies directly above (in the $B^\fourth$ axis) the QNQ
space. This is due to the onset of a neutron phase at low pressure.

\figure{Parameter space for QNQ, and QNQN stars. \label{wir3}}

\<The last space shown is for Q stars which lie below the QN quark parameter
space in Fig. \ref{wir0}. This is also due to a high pressure phase transition
to N matter at the stellar core.

\figure{Parameter space for  Q stars. \label{wir10}}

\section{Conclusions}

There is a fairly large region for Q, N, QN, and NQ stars, with neutron (N)
stars dominating all the other classes. Hybrid stars (QN) with neutron
exteriors
are the second largest space and dominate pure quark stars. Surprisingly,
NQ stars occupy a larger part of the quark parameter space than pure quark
stars as well. Not unexpectedly, the more exotic stellar classes QNQN, QNQ, and
NQN, take up less than 2\% of the total quark parameter space.

  Obervational testing may be possible. At low mass, one may be able to
distinguish between a star with a neutron matter or quark
matter at low density via observations of coalescening neutron and black
hole binaries \cite{cutler92}. These observations are unlikely to be useful
at large mass binaries where the all sequences have a similar form
(see Fig. \ref{rm}). On the otherhand, once masses and radii data for stellar
sequences are available, one may be able to determine the equation of state
for dense matter \cite{lindblom92}.

This calculation can be improved in several ways. First,
this calculation uses a value of $P_{max}$ which is not correlated
to the onset of instability directly. One could improve this calculation
by calculating $P_{critical}$ for each value of the quark parameter space
and use this value as $P_{max}$

Secondly, any particular class which is non trivial
contains all the sub-classed that can be obtained by removing the core.
A QN class also contain a pure N class. QNQN contains NQN, QN and N classs.
Thus, a more accurate ``measure" for this space is needed. One possible
solution
is to take equal units of the mass for each sequence and calculate a point
in the quark parameter space. This can be accomplished by generating sequences
of stars which differ by some constant mass, assigning a value to each star.
Obviously, both of these considerations can be integrated into the program
simultaneously.

  Finally, one could impose more restrictive values of the quark
parameters and recalculate the quark parameter space.
It could be that $\alpha_c\leq 1.0$ is the only valid
regime in the quark parameter space, and an $m_s$ which reflects experiment
should be used.

\acknowledgments
 I am grateful John L. Friedman for his many suggestions and insights.
I would also like thank to Robert R. Caldwell, Leonard Parker, A. V. Olinto
and F. Grassi for their comments and suggestions.
Computer time by National Center for Supercomputing Applications.
This work was partially supported by NFS grant No. PHY-91-05935 and a US
Department of Education Fellowship.

\appendix{Thermodynamic Potentials}

The formulas for the thermodynamic potentials below are as in Alcoch et. al.
\cite{alcock86}.

\beg\zW_u = -{\zm_u^4 \over 4 \zp^2}\left(1 - {2\za_c \over \zp}\right),\eb
\beg\zW_d = -{\zm_d^4 \over 4 \zp^2}\left(1 - {2\za_c \over \zp}\right),\eb
\beg\zW_s =-{1 \over 4 \zp^2}\left[
     \begin{array}{cccc}
     &\zm A \left(\zm^2 - \ahalf5 m_s^2\right)
           + \ahalf3 m_s^4 \ln(F) \cr
     &-{2\za_c \over \zp}
        \left\{3[\zm A - m_s^2 \ln(E)]^2 -
      2A^4 -3m_s^4 \ln^2({m_s\over\zm  }) \right.                        \cr
     &+ \left. 6\ln({\zr_r\over\zm})[\zm  m_s^2 A - m_s^4 \ln(F)] \right\} \cr
  \end{array} \right] \eb

\<and
\beg\zW_e = - {\zm_e^4\over 12\zp^2} , \eb
\<where $A=(\zm^2-m_s^2)^\half$, $F={\zm+A\over m_s}$, and $E={\zm+A\over\zm}$
{}.
$\rho_r$ is taken to be $313 MeV$ as in \cite{alcock86}.
% \label{app:subsec}
%%%%%%%%%%%%%%%%%%%%%%%%%%%%%%%%%%%%%%%%%%%%%%%%%%%%%%%%%%%%%%%%%%%%%%%%%%%%%%%
\input bibitem.tex
%%%%%%%%%%%%%%%%%%%%%%%%%%%%%%%%%%%%%%%%%%%%%%%%%%%%%%%%%%%%%%%%%%%%%%%%%%%%%%%
\end{document}